\documentclass[structabstract,onecolumn]{aa}

\usepackage{graphicx}
\usepackage{txfonts}
\usepackage{natbib}

\newcommand{\half}{\ensuremath{\frac{1}{2}}}

\newcommand{\hatbf}[1]{\ensuremath{\mathbf{\hat{#1}}}}

\newcommand{\matbf}[1]{\ensuremath{\mathbf{\stackrel{\leftrightarrow}{#1}}}}
\newcommand{\vecbf}[1]{\ensuremath{\mathbf{\stackrel{\rightarrow}{#1}}}}

\newcounter{cureqno}
\newenvironment{mathletters}{
 \refstepcounter{equation}
 \setcounter{cureqno}{\value{equation}}
 \edef\@tempa{\theequation}
 \expandafter\def
 \expandafter\theequation
 \expandafter{\@tempa\alph{equation}}
 \setcounter{equation}{0}
}{
 \setcounter{equation}{\value{cureqno}}
}

\bibpunct[; ]{(}{)}{;}{a}{}{,}

\begin{document}

\title{Combined calculi for photon orbital and spin angular momenta}

\author{N.M. Elias II\inst{1,2}}

\institute{
$^{1}$Zentrum f{\"u}r Astronomie der Universit{\"a}t Heidelberg,
Landessternwarte;
K{\"o}nigstuhl 12;
69117 Heidelberg, Germany \\
$^{2}$OAM Solutions, LLC;
9300 Stardust Trail;
Flagstaff, AZ 86004 USA
}

\date{Received December 10, 2013; Accepted XXXXXXXX xx, 2014}

\abstract
{Wavelength, photon spin angular momentum (PSAM), and photon orbital angular
momentum (POAM), completely describe the state of a photon or an electric field
(an ensemble of photons).  Wavelength relates directly to energy and linear
momentum, the corresponding kinetic quantities.  PSAM and POAM, themselves
kinetic quantities, are colloquially known as polarization and optical vortices,
respectively.  Astrophysical sources emit photons that carry this information.}
{PSAM characteristics of an electric field (intensity) are compactly described
by the Jones (Stokes/Mueller) calculus.  Similarly, I created calculi to
represent POAM characteristics of electric fields and intensities in an
astrophysical context.  Adding wavelength dependence to all of these calculi is
trivial.  The next logical steps are to 1) form photon total angular momentum
(PTAM = POAM + PSAM) calculi; 2) prove their validity using operators and
expectation values; and 3) show that instrumental PSAM can affect measured POAM
values for certain types electric fields.}
{I derive the PTAM calculi of electric fields and intensities by combining the
POAM and PSAM calculi.  I show how these quantities propagate from celestial
sphere to image plane.  I also form the PTAM operator (the sum of the POAM and
PSAM operators), with and without instrumental PSAM, and calculate the
corresponding expectation values.}
{Apart from the vector, matrix, dot product, and direct product symbols, the
PTAM and POAM calculi appear superficially identical.  I provide tables with all
possible forms of PTAM calculi.  I prove that PTAM expectation values are
correct for instruments with and without instrumental PSAM.  I also show that
POAM measurements of ``unfactored'' PTAM electric fields passing through
non-zero instrumental circular PSAM can be biased.}
{The combined PTAM calculi provide insight into how to mathematically model
PTAM sources and calibrate POAM- and PSAM- induced POAM measurement errors.}


\maketitle


\section{Introduction} \label{Sec:Introduction}

\citet{Elias1} derived propagation calculi to describe astronomical photon
orbital angular momentum (POAM; colloquially known as optical vortices).  He
employed a semi-classical/semi-quantum framework where electric fields are
analogous to photon wave functions and intensities are analogous to
probabilities.  These calculi link POAM quantities on the celestial sphere to
POAM quantities at instrument backends.  He tacitly assumes that the electric
fields on the celestial sphere are spatially uncorrelated (the ``Standard
Astronomical Assumption,'' or SAA).  \citet{Elias2} used these calculi to
describe POAM and torque metrics for single telescopes and interferometers.

Like most other workers in the POAM field, \citet{Elias1,Elias2} dealt only with
optical systems that ignored photon spin angular momentum (PSAM; colloquially
known as polarization), in order to simplify calculations.  Since POAM and PSAM
are complementary properties that will eventually be measured simultaneously,
combined calculi are required for modeling source and instrument behavior.
Failing to take non-zero PSAM into account will yield incorrect POAM values
under certain conditions.

The goals and results of this work are multiple.  First, I present the simplest
and most general photon total angular momentum (PTAM = POAM + PSAM) electric
field forms.  Second, I combine the POAM \citep{Elias1} and PSAM propagation
calculi to create the PTAM propagation calculi.  Third, I create the POAM, PSAM,
and PTAM operators and calculate the corresponding expectation values for
perfect and imperfect instruments.  Last, I show that POAM measurements can be
biased when unfactored electric fields pass through non-zero instrumental
circular PSAM.

\section{Electric Fields} \label{Sec:E_PTAM}

\citet{Elias1,Elias2} treated the electric field as a scalar quantity when he
constructed the POAM state expansions
\begin{equation} \label{Eq:E_POAM}
E(\vecbf{H};t) \, = \,
  \sum_{m=-\infty}^{\infty} E_{m}(H;t) \, e^{jm\chi}
~~~~ \stackrel{\mathcal{F}}{\Longleftrightarrow} ~~~~
E_{m}(H;t) \, = \,
  \frac{1}{2\pi} \int_{0}^{2\pi} d\chi \, e^{-jm\chi} \, E(\vecbf{H};t) \, ,
\end{equation}
where $\vecbf{H}$ $=$ $(H\cos{\chi},H\sin{\chi})$ is the vector in a plane
(e.g., celestial sphere, image plane, etc.), $t$ is time, the $m$ are the POAM
quantum numbers ($-\infty$ $\leq$ $m$ $\leq$ $\infty$), and the $E_{m}(H;t)$ are
the POAM states (azimuthal Fourier components) for each radius (perpendicular to
the propagation direction).  An azimuthal Fourier series is performed for each
$H$ and $t$.

The standard way of describing the PSAM behavior of an electric field is the
Jones vector
\begin{equation} \label{Eq:JonesV}
\vecbf{E}(\vecbf{H};t) \, = \,
  \left[
    \begin{array}{c}
      E_{R}(\vecbf{H};t) \\
      E_{L}(\vecbf{H};t)
    \end{array}
  \right] \, ,
\end{equation}
where $E_{R}(\vecbf{H};t)$ and $E_{L}(\vecbf{H};t)$ are the right-circular and
left-circular components.  The circular PSAM basis is ideal for the subsequent
analyses of this paper.

A ``factored'' electric field, where the POAM and PSAM parts are separate
factors, is the simplest and likely the most well known PTAM form
\begin{mathletters} 
\begin{equation} \label{Eq:E_PTAM_simple}
\vecbf{E}(\vecbf{H};t) \, = \,
\vecbf{\epsilon}_{POAM}(\vecbf{H};t) ~ \epsilon_{PSAM}(\vecbf{H};t) \, = \,
\left[
  \begin{array}{c}
    \epsilon_{R}(\vecbf{H};t) \\
    \epsilon_{L}(\vecbf{H};t)
  \end{array}
\right] ~
\sum_{m=-\infty}^{\infty} \epsilon_{M}(H;t) \, e^{j m \chi} \, .
\end{equation}
The most general ``unfactored'' PTAM form, on the other hand, comes from
independently expanding each PSAM component into POAM components, or
\begin{equation} \label{Eq:E_PTAM}
\vecbf{E}(\vecbf{H};t) \, = \,
\sum_{m=-\infty}^{\infty} \vecbf{E}_{m}(H;t) \, e^{jm \chi} \, = \,
  \sum_{m=-\infty}^{\infty} \left[
    \begin{array}{c}
      E_{R,m}(H;t) \\
      E_{L,m}(H;t)
    \end{array}
  \right] \, e^{jm \chi} \, = \,
\left[
  \begin{array}{c}
    \sum_{m=-\infty}^{\infty} E_{R,m}(H;t) \, e^{jm \chi} \\
    \sum_{m=-\infty}^{\infty} E_{L,m}(H;t) \, e^{jm \chi} \\
  \end{array}
\right] \, .
\end{equation}
Although POAM and PSAM appear completely intertwined, these expressions work
with the PTAM calculi (Section \ref{Sec:Calculi}) and lead to the correct
operators and expectation values (Sections \ref{Sec:ExpectationValues} and
\ref{Sec:SimpleExample}).  This type of electric field can be prepared in the
laboratory.  At present there are no known astrophysical mechanisms that
generate unfactored PTAM, but I use this form anyway for the sake of
mathematical completeness and in the event that such mechanisms will be found.
\end{mathletters}

\section{PTAM Calculi} \label{Sec:Calculi}

\citet{Elias1} created POAM propagation calculi for electric fields and
intensities using SAA.  He also treated the electric fields as scalars, ignoring
PSAM.  Their time-averaged square magnitudes are intensities, which are
analogous to Stokes I.

In this section, I combine the \citet{Elias1} POAM calculi with the
electric-field PSAM calculi of \citet{Jones} and the intensity PSAM calculi of
\citet{Stokes} and \citet{Mueller} to form the PTAM calculi.  I also employ the
mathematics of \citet{Schmeider} and \citet{Barakat} (hereafter, SB) to more
easily link Jones vectors, Jones matrices, Stokes vectors, and Mueller matrices
via direct products and coherence matrices (assuming no system depolarization).

\subsection{POAM Correlations} \label{SSec:Correlations}

According to SB, the coherence vector is the direct product of the electric
field from Equation \ref{Eq:JonesV}
\begin{equation} \label{Eq:C}
\vecbf{C}(\vecbf{H}) \, = \,
  \left<
    \frac{1}{2} \vecbf{E}(\vecbf{H};t) \otimes \vecbf{E}^{*}(\vecbf{H};t)
  \right> \, = \,
\left[
  \begin{array}{c}
    \left< \frac{1}{2} E_{R}(\vecbf{H};t) E^{*}_{R}(\vecbf{H};t) \right> \\
    \left< \frac{1}{2} E_{R}(\vecbf{H};t) E^{*}_{L}(\vecbf{H};t) \right> \\
    \left< \frac{1}{2} E_{L}(\vecbf{H};t) E^{*}_{R}(\vecbf{H};t) \right> \\
    \left< \frac{1}{2} E_{L}(\vecbf{H};t) E^{*}_{L}(\vecbf{H};t) \right>
  \end{array}
\right] \, = \,
\left[
  \begin{array}{c}
    C_{R,R}(\vecbf{H}) \\
    C_{R,L}(\vecbf{H}) \\
    C_{L,R}(\vecbf{H}) \\
    C_{L,L}(\vecbf{H})
  \end{array}
\right] \, = \,
\left[
  \begin{array}{c}
    C_{R,R}(\vecbf{H}) \\
    C_{R,L}(\vecbf{H}) \\
    C^{*}_{R,L}(\vecbf{H}) \\
    C_{L,L}(\vecbf{H})
  \end{array}
\right] \, ,
\end{equation}
where $\otimes$ is the direct (outer) product, and $\left< \, \right>$ is the
time average.  If I substitute Equation \ref{Eq:E_PTAM} (instead of Equation
\ref{Eq:JonesV}) into Equation \ref{Eq:C}, I obtain
\begin{mathletters} \label{Eq:C_Expansion}
\begin{equation} \label{Eq:C_PTAM}
\vecbf{C}(\vecbf{H}) \, = \,
\sum_{m=-\infty}^{\infty} \sum_{n=-\infty}^{\infty}
  \left<
    \frac{1}{2} \vecbf{E}_{m}(H;t) \otimes \vecbf{E}^{*}_{n}(H;t)
  \right> e^{j(m-n)\chi} \, = \,
\sum_{m=-\infty}^{\infty} \sum_{n=-\infty}^{\infty}
  \vecbf{C}_{m,n}(H) \, e^{j(m-n)\chi} \nonumber \, ,
\end{equation}
where
\begin{equation} \label{C_mn}
\vecbf{C}_{m,n}(H) \, = \,
  \left[
    \begin{array}{c}
      \left<\frac{1}{2} E_{R,m}(H;t) E^{*}_{R,n}(H;t)\right> \\
      \left<\frac{1}{2} E_{R,m}(H;t) E^{*}_{L,n}(H;t)\right> \\
      \left<\frac{1}{2} E_{L,m}(H;t) E^{*}_{R,n}(H;t)\right> \\
      \left<\frac{1}{2} E_{L,m}(H;t) E^{*}_{L,n}(H;t)\right>
    \end{array}
  \right] \, = \,
  \left[
    \begin{array}{c}
      C_{(R,m),(R,n)}(H) \\
      C_{(R,m),(L,n)}(H) \\
      C_{(L,m),(R,n)}(H) \\
      C_{(L,m),(L,n)}(H)
    \end{array}
  \right]
\end{equation}
is the $(m,n)^{th}$ POAM correlation of the coherence vector.  By comparing
Equations \ref{Eq:C} and \ref{Eq:C_Expansion}a-b, I find that the individual
correlations can be expanded into double sums
\end{mathletters}
\begin{mathletters} \label{Eq:C_Expansion_mn}
\begin{equation} \label{Eq:C_Rm_Rn}
C_{R,R}(\vecbf{H}) \, = \,
  \sum_{m=-\infty}^{\infty} \sum_{n=-\infty}^{\infty}
  C_{(R,m),(R,n)}(H) \, e^{j(m-n)\chi} \, ,
\end{equation}
\begin{equation} \label{Eq:C_Rm_Ln}
C_{R,L}(\vecbf{H}) \, = \,
  \sum_{m=-\infty}^{\infty} \sum_{n=-\infty}^{\infty}
  C_{(R,m),(L,n)}(H) \, e^{j(m-n)\chi} \, ,
\end{equation}
\begin{equation} \label{Eq:C_Lm_Rn}
C_{L,R}(\vecbf{H}) \, = \,
  \sum_{m=-\infty}^{\infty} \sum_{n=-\infty}^{\infty}
  C_{(L,m),(R,n)}(H) \, e^{j(m-n)\chi} \, ,
\end{equation}
\noindent and
\begin{equation} \label{Eq:C_Lm_Ln}
C_{L,L}(\vecbf{H}) \, = \,
  \sum_{m=-\infty}^{\infty} \sum_{n=-\infty}^{\infty}
  C_{(L,m),(L,n)}(H) \, e^{j(m-n)\chi} \, .
\end{equation}
The $\vecbf{C}_{m,n}(H)$ have a similar form to and the identical units as the
$I_{m,n}(H)$ POAM correlations defined by \citet{Elias1}.
\end{mathletters}

Coherence vectors are used mostly by engineers because some instruments, such as
radio interferometers, employ right- and left- circular feeds.  Astronomers
prefer Stokes vectors because their components represent the total intensity and
the polarization parameters required for scientific analysis.  The Stokes vector
is related to the coherence vector via a simple matrix transformation (SB)
\begin{mathletters}
\begin{equation} \label{Eq:S}
\vecbf{S}(\vecbf{H}) \, = \,
\matbf{T} \, \cdot \, \vecbf{C}(\vecbf{H}) \, = \,
\left[
  \begin{array}{c}
    C_{R,R}(\vecbf{H}) + C_{L,L}(\vecbf{H}) \\
    C_{R,L}(\vecbf{H}) + C^{*}_{R,L}(\vecbf{H}) \\
    -j \left[ C_{R,L}(\vecbf{H}) - C^{*}_{R,L}(\vecbf{H}) \right] \\
    C_{R,R}(\vecbf{H}) - C_{L,L}(\vecbf{H})
  \end{array}
\right] \, = \,
\left[
  \begin{array}{c}
    C_{R,R}(\vecbf{H}) + C_{L,L}(\vecbf{H}) \\
    2 \, \mathrm{Re} \, C_{R,L}(\vecbf{H}) \\
    2 \, \mathrm{Im} \, C_{R,L}(\vecbf{H}) \\
    C_{R,R}(\vecbf{H}) - C_{L,L}(\vecbf{H})
  \end{array}
\right] \, = \,
\left[
  \begin{array}{c}
    I(\vecbf{H}) \\
    Q(\vecbf{H}) \\
    U(\vecbf{H}) \\
    V(\vecbf{H})
  \end{array}
\right] \, ,
\end{equation}
where $\cdot$ is the dot (inner) product, and
\begin{equation} \label{Eq:T}
\matbf{T} \, = \,
  \left[
    \begin{array}{cccc}
      1 & 0 & 0 & 1 \\
      0 & 1 & 1 & 0 \\
      0 & -j & j & 0 \\
      1 & 0 & 0 & -1
    \end{array}
  \right]
\end{equation}
is the coherence-to-Stokes transformation matrix in the circular basis.  If I
substitute Equations \ref{Eq:C_Expansion}a-b into Equation \ref{Eq:S}, I obtain
\end{mathletters}
\begin{mathletters} \label{Eq:S_Expansion}
\begin{equation} \label{Eq:S_PTAM}
\vecbf{S}(\vecbf{H}) \, = \,
\sum_{m=-\infty}^{\infty} \sum_{n=-\infty}^{\infty}
  \left[ \matbf{T} \cdot \vecbf{C}_{m,n}(H) \right] e^{j(m-n)\chi} \, = \,
\sum_{m=-\infty}^{\infty} \sum_{n=-\infty}^{\infty}
  \vecbf{S}_{m,n}(H) \, e^{j(m-n)\chi} \, ,
\end{equation}
where
\begin{equation} \label{Eq:S_mn}
\vecbf{S}_{m,n}(H) \, = \,
  \left[
    \begin{array}{c}
      C_{(R,m),(R,n)}(H) + C_{(L,m),(L,n)}(H) \\
      C_{(R,m),(L,n)}(H) + C_{(L,m),(R,n)}(H) \\
      -j \left[ C_{(R,m),(L,n)}(H) - C_{(L,m),(R,n)}(H) \right] \\
      C_{(R,m),(R,n)}(H) - C_{(L,m),(L,n)}(H)
  \end{array}
  \right] \, = \,
  \left[
    \begin{array}{c}
      I_{m,n}(H) \\
      Q_{m,n}(H) \\
      U_{m,n}(H) \\
      V_{m,n}(H)
    \end{array}
  \right]
\end{equation}
is the $(m,n)^{th}$ POAM correlation of the Stokes vector.  By comparing
Equations \ref{Eq:S} and \ref{Eq:S_Expansion}a-b, I find that the individual
Stokes parameters can be expanded into double sums
\end{mathletters}
\begin{mathletters}
\begin{equation} \label{Eq:I_mn}
I(\vecbf{H}) \, = \,
  \sum_{m=-\infty}^{\infty} \sum_{n=-\infty}^{\infty}
  I_{m,n}(H) \, e^{j(m-n)\chi} \, ,
\end{equation}
\begin{equation} \label{Eq:Q_mn}
Q(\vecbf{H}) \, = \,
  \sum_{m=-\infty}^{\infty} \sum_{n=-\infty}^{\infty}
  Q_{m,n}(H) \, e^{j(m-n)\chi} \, ,
\end{equation}
\begin{equation} \label{Eq:U_mn}
U(\vecbf{H}) \, = \,
  \sum_{m=-\infty}^{\infty} \sum_{n=-\infty}^{\infty}
  U_{m,n}(H) \, e^{j(m-n)\chi} \, ,
\end{equation}
\noindent and
\begin{equation} \label{Eq:V_mn}
V(\vecbf{H}) \, = \,
  \sum_{m=-\infty}^{\infty} \sum_{n=-\infty}^{\infty}
  V_{m,n}(H) \, e^{j(m-n)\chi} \, .
\end{equation}
The $\vecbf{S}_{m,n}(H)$ have a similar form to and the same units as the
$I_{m,n}(H)$ POAM correlations defined by \citet{Elias1}.  As a matter of fact,
Equation \ref{Eq:I_mn} is identical to the expansion derived by \citet{Elias1}
using scalar electric fields.
\end{mathletters}

The Stokes- Q, U, and V expansions are unnecessary, so I rewrite the Stokes
vector as
\begin{equation} \label{Eq:S_PTAM_CS2}
\vecbf{S}(\vecbf{H}) \, = \,
  \left[
    \begin{array}{c}
      \sum_{m=-\infty}^{\infty} \sum_{n=-\infty}^{\infty}
        I_{m,n}(H) \, e^{j(m-n)\chi} \\
        Q(\vecbf{H}) \\
        U(\vecbf{H}) \\
        V(\vecbf{H})
    \end{array}
  \right] \, .
\end{equation}
This PTAM form maintains both the POAM and PSAM information while minimizing
complications.

\subsection{POAM Rancors} \label{SSec:Rancors}

\citet{Elias1} defined a quantity called rancor, which is the azimuthal Fourier
series versus radius of the intensity
\begin{mathletters}
\begin{equation} \label{Eq:Rancor1}
I(\vecbf{H}) \, = \,
  \sum_{m=-\infty}^{\infty} \mathcal{I}_{m}(H) \, e^{jm\chi}
~~~~ \stackrel{\mathcal{F}}{\Longleftrightarrow} ~~~~
\mathcal{I}_{m}(H) \, = \,
  \frac{1}{2\pi} \int_{0}^{2\pi} d\chi \, e^{-jm\chi} \, I(\vecbf{H}) \, ,
\end{equation}
where $\mathcal{I}_{m}(H)$ is the $m^{th}$ POAM rancor.  This quantity is
interesting because it identical to the infinite sum over a subset of POAM
correlations
\begin{equation} \label{Eq:Rancor2}
\mathcal{I}_{m}(H) \, = \, \sum_{k=-\infty}^{\infty} I_{k,k-m}(H) \, .
\end{equation}
Rancors, which may be easier to determine in some cases, contain a limited
amount of POAM information.  As an analogy, I point out that squared visbilities
and closure phases in optical interferometry can provide important physical data
about astronomical sources, in spite of the fact that they contain less
information than complex visibilities.
\end{mathletters}

In Section \ref{SSec:Correlations}, I combined POAM correlations with PSAM
Stokes vectors.  Since rancors can be expressed in terms of correlations, it
follows that all intensity formulae in Section \ref{SSec:Correlations} can be
written in terms of rancors.  I will not list all possible expressions here,
since those expansions are identical to those in
\citet[Sections 4-5, Appendix C]{Elias1} apart
from the fact that scalar quantities are replaced by vectors and matrices and
scalar products are replaced by dot and direct products.

\subsection{Propagating POAM Quantities} \label{SSec:Propagating}

\citet{Elias1} derived scalar electric-field and intensity calculi for
propagating POAM from celestial sphere to image plane and listed them in several
tables.  He employed system forms and SAA.  In this section, I extend these
expressions to combine POAM and PSAM propagation calculi, thus creating PTAM
propagation calculi.

Consider the system form for propagation of the scalar electric field from
celestial sphere to image plane
\begin{mathletters}
\begin{equation} \label{Eq:EPropScalar}
E(\vecbf{\Omega}^{\prime};t) \, = \,
  \int d^{2}\Omega \, D(\vecbf{\Omega}^{\prime},\vecbf{\Omega}) \,
  E(\vecbf{\Omega};t) \, ,
\end{equation}
where $\vecbf{\Omega}^{\prime}$ $=$
$(\rho^{\prime} \cos{\phi^{\prime}},\rho^{\prime} \sin{\phi^{\prime}})$ is the
coordinate in the image plane, $\vecbf{\Omega}$ $=$
$(\rho \cos{\phi}, \rho \sin{\phi})$ is the coordinate on the celestial sphere,
\begin{equation} \label{Eq:DiffractionFunction}
D(\vecbf{\Omega}^{\prime},\vecbf{\Omega}) \, = \,
  \int d^{2}r \,
  e^{
    j 2\pi \vecbf{r} \cdot \left(
      \vecbf{\Omega}^{\prime} - \vecbf{\Omega}
    \right)
  } \, D(\vecbf{r})
\end{equation}
is the diffraction function, $e^{ j 2\pi \vecbf{r} \cdot \left(
\vecbf{\Omega}^{\prime} - \vecbf{\Omega} \right)}$ is the Fraunhofer propagator
(it can be replaced with the Fresnel propagator), $\vecbf{r}$ $=$ $(r
\cos{\psi},r \sin{\psi})$ is the coordinate in the pupil plane normalized by
wavelength, and D(\vecbf{r}) is the pupil function which describes the telescope
aberrations, atmospheric turbulence, etc.  If these scalar electric fields are
changed to 2x1 Jones vectors, the diffraction function must become a 2x2 Jones
matrix
\begin{equation} \label{Eq:EPropPSAM}
\vecbf{E}(\vecbf{\Omega}^{\prime};t) \, = \,
  \int d^{2}\Omega \, \matbf{D}(\vecbf{\Omega}^{\prime},\vecbf{\Omega})
  \, \cdot \, \vecbf{E}(\vecbf{\Omega};t) \, .
\end{equation}
In principle, the scalar and matrix diffraction functions can also be functions
of time, although their variability time scales are much slower than those of
the electric fields.
\end{mathletters}

If the Jones vector components are expanded into independent POAM states
(unfactored form, Equation \ref{Eq:E_PTAM}), Equation \ref{Eq:EPropPSAM} becomes
the PTAM state expansion
\begin{mathletters}
\begin{equation} \label{Eq:EPropPTAM}
\vecbf{E}(\vecbf{\Omega}^{\prime};t) \, = \,
  \sum_{p=-\infty}^{\infty} \vecbf{E}_{p}(\rho^{\prime};t) \,
  e^{j p \phi^{\prime}} \, = \,
\sum_{p=-\infty}^{\infty} \left[
  \sum_{m=-\infty}^{\infty} 2\pi \int_{0}^{\infty} d\rho \, \rho \,
  \matbf{D}^{-m}_{p}(\rho^{\prime},\rho) \, \cdot \, \vecbf{E}_{m}(\rho;t)
\right] \, e^{j p \phi^{\prime}} \, ,
\end{equation}
where
\begin{equation} \label{Eq:EPTAMImage}
\vecbf{E}_{p}(\rho^{\prime};t) \, = \,
  \frac{1}{2\pi} \int_{0}^{2\pi} d\phi^{\prime} \, e^{-j p \phi^{\prime}} \,
  \vecbf{E}(\vecbf{\Omega}^{\prime};t)
\end{equation}
is the output POAM state $p$,
\begin{equation} \label{Eq:EPTAMCS}
\vecbf{E}_{m}(\rho;t) \, = \,
  \frac{1}{2\pi} \int_{0}^{2\pi} d\phi \, e^{-j m \phi} \,
  \vecbf{E}(\vecbf{\Omega};t)
\end{equation}
is the input POAM state $m$, and
\begin{equation} \label{Eq:DiffFuncPTAM}
\matbf{D}^{-m}_{p}(\rho^{\prime},\rho) \, = \,
  \frac{1}{2\pi} \int_{0}^{2\pi} d\phi^{\prime} \, e^{-j p \phi^{\prime}} \,
  \frac{1}{2\pi} \int_{0}^{2\pi} d\phi \, e^{j m \phi} \,
  \matbf{D}(\vecbf{\Omega}^{\prime},\vecbf{\Omega})
\end{equation}
is the diffraction function gain between output POAM state $p$ and the input
POAM state $m$.  I summarize all PTAM electric field expansions in Tables
\ref{Tab:ExpE} and \ref{Tab:ExpD}.
$\vecbf{\mathfrak{E}}_{m}(\vecbf{\Omega}^{\prime};\vecbf{a},t)$ is not a true
PTAM state, which means that the input expansion is of limited use but included
for the sake of completeness.
\end{mathletters}

The intensity is the squared magnitude of the electric field.  Using SAA and
Equation \ref{Eq:EPropScalar}, the scalar intensity becomes
\begin{mathletters}
\begin{equation} \label{Eq:IPropScalar}
I(\vecbf{\Omega}^{\prime}) \, = \,
\left< \half \left|E(\vecbf{\Omega}^{\prime};t)\right|^{2} \right> \, = \,
  \int d^{2}\Omega \, P(\vecbf{\Omega}^{\prime},\vecbf{\Omega}) \,
  I(\vecbf{\Omega}) \, ,
\end{equation}
where $P(\vecbf{\Omega}^{\prime},\vecbf{\Omega})$ $=$ $\left|
D(\vecbf{\Omega}^{\prime},\vecbf{\Omega}) \right|^{2}$ is the point-spread
function (PSF), and $I(\vecbf{\Omega})$ $=$
$\left<\half\left|E(\vecbf{\Omega};t)\right|^{2}\right>$.  SAA collapses one of
the integrals over the celestial sphere.  If I employ Equations \ref{Eq:C},
\ref{Eq:S}, and \ref{Eq:EPropPSAM} as well as SAA, the scalar Equation
\ref{Eq:IPropScalar} becomes the vector equation
\begin{eqnarray} \label{Eq:IPropPSAM}
\vecbf{S}(\vecbf{\Omega}^{\prime}) &=
  &\matbf{T} \cdot \left<
    \frac{1}{2} \vecbf{E}(\vecbf{\Omega}^{\prime};t) \otimes
    \vecbf{E}^{*}(\vecbf{\Omega}^{\prime};t)
  \right> \, = \,
  \int d^{2}\Omega \, \left\{
    \matbf{T} \cdot \left[
      \matbf{D}(\vecbf{\Omega}^{\prime},\vecbf{\Omega}) \otimes
      \matbf{D}^{*}(\vecbf{\Omega}^{\prime},\vecbf{\Omega})
    \right] \cdot \matbf{T}^{-1}
  \right\} \cdot \left\{
    \matbf{T} \cdot \left<
      \frac{1}{2} \vecbf{E}(\vecbf{\Omega};t) \otimes
      \vecbf{E}^{*}(\vecbf{\Omega};t)
    \right>
  \right\} \nonumber \\ &=
  &\int d^{2}\Omega \, \matbf{P}(\vecbf{\Omega}^{\prime},\vecbf{\Omega})
  \, \cdot \, \vecbf{S}(\vecbf{\Omega}) \, .
\end{eqnarray}
The point spread function is now a 4x4 Mueller matrix.  I summarize all PTAM
intensity expansions in Tables \ref{Tab:ExpS} and \ref{Tab:ExpP}. 
$\vecbf{\mathfrak{S}}_{m,n}(\vecbf{\Omega}^{\prime};\vecbf{a})$ and
$\vecbf{\mathfrak{S}}_{m}(\vecbf{\Omega}^{\prime};\vecbf{a})$ are not true PTAM
quantities, which means that the input expansions are of limited use but
included for the sake of completeness.  Also, note that the intensity equations
are cannot be derived from the electric field equations when a system has
depolarization (Mueller matrices cannot be uniquely determined from Jones
matrices).
\end{mathletters}

Now consider the Stokes-I parameter the image plane
\begin{eqnarray} \label{Eq:StokesImage}
I(\vecbf{\Omega}^{\prime}) &=
&\hatbf{d}^{T} \cdot \vecbf{S}(\vecbf{\Omega}^{\prime}) \, = \,
\int d^{2}\Omega \, \hatbf{d}^{T} \, \cdot \,
  \matbf{P}(\vecbf{\Omega}^{\prime},\vecbf{\Omega}) \, \cdot \,
  \vecbf{S}(\vecbf{\Omega}) \nonumber \\ &=
&\int d^{2}\Omega \,
  \mathbf{P}^{I,I}(\vecbf{\Omega}^{\prime},\vecbf{\Omega}) \, I(\vecbf{\Omega})
  \, + \, \int d^{2}\Omega \,
  \mathbf{P}^{I,Q}(\vecbf{\Omega}^{\prime},\vecbf{\Omega}) \, Q(\vecbf{\Omega})
  \, + \, \int d^{2}\Omega \,
  \mathbf{P}^{I,U}(\vecbf{\Omega}^{\prime},\vecbf{\Omega}) \, U(\vecbf{\Omega})
  \, + \, \int d^{2}\Omega \,
  \mathbf{P}^{I,V}(\vecbf{\Omega}^{\prime},\vecbf{\Omega}) \,
  V(\vecbf{\Omega}) \, ,
\end{eqnarray}
where $\hatbf{d}^{T} = [1,0,0,0]$ is the detector operator, and the
$\mathbf{P}^{I,x}(\vecbf{\Omega}^{\prime},\vecbf{\Omega})$ are the elements of
the top row of the Mueller matrix PSF.  In Sections \ref{SSec:Correlations} and
\ref{SSec:Rancors}, I point out that only the Stokes-I parameter must be
expanded in terms of POAM correlations and rancors, even though the complete
derivations involve POAM-like expansions of the other Stokes parameters.
Similarly, only the upper-left element of the Mueller matrix
$\mathbf{P}^{I,I}(\vecbf{\Omega}^{\prime},\vecbf{\Omega})$ must be expanded in
terms of POAM correlations or rancors.  Equation \ref{Eq:StokesImage} indicates
that non-zero Stokes- Q, U, and V terms could introduce measurement biases
which must be calibrated when measuring the POAM of the Stokes-I parameter.  I
present a simple example in Section \ref{Sec:SimpleExample} using operators and
expectation values.

\section{Operators and Expectation Values} \label{Sec:ExpectationValues}

Expectation values are specific quantities that can be measured by instruments.
In this section, I: 1) define the POAM, PSAM, and PTAM operators; 2) derive the
corresponding expectation values; 3) show how the operators and expectation
values are modified by imperfect instruments.

\subsection{Perfect Instrument} \label{SSec:PerfectInstrument}

In the paraxial case, the quantum mechanical POAM operator along the $+z$
propagation axis is
\begin{equation} \label{Eq:Lop}
L_{Z}(\vecbf{H}) \, \rightarrow \, L_{Z}(\chi) \, = \,
  j \hbar \, \frac{\partial}{\partial \chi} \, ,
\end{equation}
where $j$ $=$ $\sqrt{-1}$, and $\hbar$ is Planck's constant $h$ divided by
$2\pi$.  The POAM expectation value is measured when this operator is
applied to the scalar electric field
\begin{mathletters}
\begin{equation} \label{Eq:Lexp}
\hat{L}_{Z} \, = \,
  \frac{1}{I_{s}} \int d^{2}H
  \left<
    \frac{1}{2} E(\vecbf{H};t) \, L_{Z}(\vecbf{H}) \, E^{*}(\vecbf{H};t)
  \right> \, ,
\end{equation}
where
\begin{equation} \label{Eq:I_integrated_scalar}
I_{s} \, = \,
  \int d^{2}H \, I_{s}(\vecbf{H}) \, = \,
\int d^{2}H \, \left<
  \frac{1}{2} E(\vecbf{H};t) \, E^{*}(\vecbf{H};t)
\right> \, = \,
\int d^{2}H \, \left< \frac{1}{2} \left| E(\vecbf{H};t) \right|^{2} \right>
\end{equation}
is the integrated intensity of the scalar electric field.  The numerator is a
quantum-mechanics-like product of states and matrix elements, and the
denominator is the normalization.  Substituting Equations \ref{Eq:E_POAM} and
\ref{Eq:Lop} into Equation \ref{Eq:Lexp}, I obtain
\end{mathletters}
\begin{mathletters}
\begin{equation} \label{Eq:LexpFinal}
\hat{L}_{Z} \, = \,
  M_{eff} \hbar \, = \,
\left\{ \sum_{m=-\infty}^{\infty} m p_{m,m} \right\} \hbar \, ,
\end{equation}
where $M_{eff}$ is the effective quantum number,
\begin{equation} \label{Eq:p_mm}
p_{m,m} \, = \,
\frac{I_{m,m}}{I_{s}} \, = \,
\frac{1}{I_{s}} \, 2\pi \int_{0}^{r_{max}} dH H \, I_{m,m}(H) \, = \,
\frac{1}{I_{s}} \, 2\pi \int_{0}^{r_{max}} dH H \,
  \left< \frac{1}{2}|E_{m}(H;t)|^{2} \right>
\end{equation}
is the probability of a photon (or an enesmble of photons) being in state $m$,
$r_{max}$ is the maximum radius which contains all of the flux, and $I_{m,m}$ is
the radially integrated autocorrelation of POAM state $m$.  The expection value
is simply the effective quantum number times $\hbar$.
\end{mathletters}

Similarly, the quantum mechanical PSAM operator along the $+z$ propagation axis
is
\begin{equation} \label{Eq:Sop}
\matbf{S}_{Z}(\vecbf{H}) \, \rightarrow \,
\matbf{S}_{Z} \, = \,
  \hbar \, \matbf{\sigma}_{3} \, = \,
  \hbar \, \left[
    \begin{array}{cc}
      1 & 0 \\
      0 & -1 \\
    \end{array}
  \right] \, ,
\end{equation}
where $\matbf{\sigma}_{3}$ is the third Pauli spin matrix.  The PSAM expectation
value is measured when this operator is applied to the vector electric field
\begin{mathletters}
\begin{eqnarray} \label{Eq:Sexp}
\hat{S}_{Z} \, = \,
\frac{1}{I} \int d^{2}H
  \left<
    \frac{1}{2} \vecbf{E}^{T}(\vecbf{H};t) \cdot \matbf{S}_{Z}(\vecbf{H}) \cdot
    \vecbf{E}^{*}(\vecbf{H};t)
  \right> \, ,
\end{eqnarray}
where
\begin{equation} \label{Eq:I_integrated_vector}
I \, = \,
  \int d^{2}H \, I(\vecbf{H}) \, = \,
\int d^{2}H \, \left<
  \frac{1}{2} \vecbf{E}^{T}(\vecbf{H};t) \cdot \vecbf{E}^{*}(\vecbf{H};t)
\right> \, = \,
\int d^{2}H \, \hatbf{d}^{T} \cdot \matbf{T} \cdot \left<
  \frac{1}{2} \vecbf{E}(\vecbf{H};t) \otimes \vecbf{E}^{*}(\vecbf{H};t)
\right> \, ,
\end{equation}
is the integrated intensity of the vector electric field, and the $T$
superscript indicates the transpose.  Substituting Equations \ref{Eq:JonesV} and
\ref{Eq:Sop} into Equation \ref{Eq:Sexp}, I obtain
\end{mathletters}
\begin{mathletters}
\begin{equation} \label{Eq:SexpFinal}
\hat{S}_{Z} \, = \,
  v \hbar \, = \,
  \left\{ p_{R,R} - p_{L,L} \right\} \hbar \, ,
\end{equation}
where $v$ is the normalized Stokes-V parameter,
\begin{equation} \label{Eq:p_R}
p_{R,R} \, = \,
\frac{I_{R,R}}{I} \, = \,
\frac{1}{I} \int d^{2}H \, I_{R,R}(\vecbf{H}) \, = \,
\frac{1}{I} \int d^{2}H \, \left< \frac{1}{2}|E_{R}(\vecbf{H};t)|^{2} \right>
\end{equation}
and
\begin{equation} \label{Eq:p_L}
p_{L,L} \, = \,
\frac{I_{L,L}}{I} \, = \,
\frac{1}{I} \int d^{2}H \, I_{L,L}(\vecbf{H}) \, = \,
\frac{1}{I} \int d^{2}H \, \left< \frac{1}{2}|E_{L}(\vecbf{H};t)|^{2} \right>
\end{equation}
are the probabilities of a photon (or an ensemble of photons) being in the RCP
and LCP states, and $I_{R,R}$ and $I_{L,L}$ are the integrated autocorrelations
of the RCP and LCP states. For an unpolarized and/or linearly polarized source
$v$ $=$ $0$, which means that $p_{R,R}$ $=$ $p_{L,L}$ $=$ $\half$.  For a fully
circularly polarized source, $v$ $=$ $+1$ ($v$ $=$ $-1$), $p_{R,R}$ $=$ $1$ and
$p_{L,L}$ $=$ $0$ ($p_{R,R}$ $=$ $0$ and $p_{L,L}$ $=$ $1$).
\end{mathletters}

The PTAM expectation value is the sum of the POAM and PSAM expectation values,
or $\hat{J}_{Z}$ $=$ $\hat{L}_{Z}$ $+$ $\hat{S}_{Z}$.  The PTAM expectation
value can be measured directly with the PTAM operator
$\matbf{J}_{Z}(\vecbf{H})$ instead, but the POAM operator must first be
converted to a matrix
\begin{equation} \label{Eq:LopVector}
L_{Z}(\vecbf{H}) \, \rightarrow \,
L_{Z}(\chi) \, \Rightarrow \,
\matbf{L}_{Z}(\vecbf{H}) \, \rightarrow \,
\matbf{L}_{Z}(\chi) \, = \,
  \matbf{\sigma}_{0} \, j \hbar \, \frac{\partial}{\partial \chi} \, ,
\end{equation}
where $\matbf{\sigma}_{0}$ $=$ $\matbf{1}$ is the zeroth Pauli spin matrix (2x2
identity matrix).  With this redefined POAM operator, the PTAM operator becomes
\begin{mathletters}
\begin{eqnarray} \label{Eq:Jop}
\matbf{J}_{Z}(\vecbf{H}) &=
&\matbf{L}_{Z}(\vecbf{H}) \, + \, \matbf{S}_{Z}(\vecbf{H}) \, \rightarrow \,
  \matbf{J}_{Z}(\chi) \, = \, \matbf{L}_{Z}(\chi) \, + \, \matbf{S}_{Z}
  \nonumber \\ &=
&\hbar \, \left[
  j \, \matbf{\sigma}_{0} \, \frac{\partial}{\partial \chi} \, + \,
  \matbf{\sigma}_{3}
\right] \, = \,
\hbar \, \left[
  \begin{array}{cc}
    j \frac{\partial}{\partial \chi} \, + \, 1 & 0 \\
    0 & j \frac{\partial}{\partial \chi} \, - \, 1 \\
  \end{array}
\right] \, .
\end{eqnarray}
Thus,
\begin{eqnarray} \label{Eq:Jexp}
\hat{J}_{Z} &=
&\hat{L}_{Z} \, + \, \hat{S}_{Z} \nonumber \\ &=
&\frac{1}{I} \int d^{2}H
  \left<
    \frac{1}{2} \vecbf{E}^{T}(\vecbf{H};t) \cdot \matbf{J}_{Z}(\vecbf{H}) \cdot
    \vecbf{E}^{*}(\vecbf{H};t)
  \right> \nonumber \\ &=
&\frac{1}{I}
  \int d^{2}H
  \left<
    \frac{1}{2} \vecbf{E}^{T}(\vecbf{H};t) \cdot \matbf{L}_{Z}(\vecbf{H}) \cdot
    \vecbf{E}^{*}(\vecbf{H};t)
  \right> +
\frac{1}{I}
  \int d^{2}H
  \left<
    \frac{1}{2} \vecbf{E}^{T}(\vecbf{H};t) \cdot \matbf{S}_{Z}(\vecbf{H}) \cdot
    \vecbf{E}^{*}(\vecbf{H};t)
  \right> \nonumber \\ &=
&\left[ M_{eff} \, + \, v \right] \, \hbar \, = \,
\left[
  \left\{ \sum_{m=-\infty}^{\infty} m p_{m,m} \right\} \, + \,
  \left\{ p_{R,R} - p_{L,L} \right\}
\right] \, \hbar \, .
\end{eqnarray}
The choice of measuring $\hat{J}_{Z}$ using separate $\matbf{L}_{Z}(\chi)$ and
$\matbf{S}_{Z}$ operators or the combined $\matbf{J}_{Z}(\chi)$ operator depends
on the application.
\end{mathletters}

\subsection{Imperfect Instrument} \label{SSec:ImperfectInstrument}

An instrument with non-zero instrumental PSAM, subject to the equations of
Section \ref{SSec:Propagating}, modifies the expectation values derived in
Section \ref{SSec:PerfectInstrument}.  For the sake of simplicity, I assume
that the circular telescope aperture is uniformly unaberrated with a non-zero
instrumental PSAM, which means that
\begin{mathletters}
\begin{equation} \label{Eq:JonesMatrixSimple}
\matbf{D}(\vecbf{r}) \, \rightarrow \,
\matbf{D} \, = \,
  \left[
    \begin{array}{cc}
      D^{A,A} & D^{A,B} \\
      D^{B,A} & D^{B,B}
    \end{array}
  \right] \, ,
\end{equation}
\begin{equation} \label{Eq:DiffFunc}
\matbf{D}(\vecbf{\Omega}^{\prime},\vecbf{\Omega}) \, = \,
  \matbf{D} ~ \pi R^{2}_{tel} ~
  \mathrm{jinc}\left(
    2\pi \, R_{tel} \left|\vecbf{\Omega}^{\prime}-\vecbf{\Omega}\right|
  \right) \, ,
\end{equation}
and
\begin{equation} \label{Eq:EDiffFunc}
\vecbf{E}(\vecbf{\Omega}^{\prime};t) \, = \,
  \int d^{2}\Omega \, \matbf{D}(\vecbf{\Omega}^{\prime},\vecbf{\Omega}) \cdot
  \vecbf{E}(\vecbf{\Omega};t) \, = \,
\matbf{D} \, \cdot \, \int d^{2}\Omega \left\{
  \pi R^{2}_{tel} ~ \mathrm{jinc}\left(
    2\pi R_{tel} \, \left|\vecbf{\Omega}^{\prime}-\vecbf{\Omega}\right|
  \right)
\right\} \, \vecbf{E}(\vecbf{\Omega};t) \, = \,
\matbf{D} \, \cdot \, \vecbf{\mathcal{E}}(\vecbf{\Omega}^{\prime};t) \, ,
\end{equation}
where $R_{tel}$ is the telescope radius in units of wavelength,
$\mathrm{jinc}\left(x\right)$ $=$ $2 J_{1}\left(x\right) / x$, and
$J_{1}\left(x\right)$ is the Bessel function of the first kind of order one.
The quantity in the curly braces approaches the Dirac delta function
$\delta(\vecbf{\Omega}^{\prime} - \vecbf{\Omega})$ when $R_{tel}$ $\rightarrow$
$\infty$.  A perfect instrument implies that $\matbf{D}$ $=$ $\eta$ $\matbf{1}$,
where $\eta$ is a complex constant ($0 < \left|\eta\right| \leq 1$).
Conversely, when $\matbf{D}$ $\neq$ $\eta$ $\matbf{1}$ the instrument mixes the
PSAM components.
\end{mathletters}

To keep the notation consistent with Section \ref{SSec:PerfectInstrument}, I
let $\vecbf{\Omega}^{\prime}$ $\rightarrow$ $\vecbf{H}$,
$\vecbf{E}(\vecbf{\Omega}^{\prime};t)$ $\rightarrow$
$\vecbf{E}^{\prime}(\vecbf{H};t)$, and
$\vecbf{\mathcal{E}}(\vecbf{\Omega}^{\prime};t)$ $\rightarrow$
$\vecbf{E}(\vecbf{H};t)$.  The PTAM expectation value for this imperfect
instrument is
\begin{mathletters}
\begin{eqnarray} \label{Eq:JopSimple}
\hat{J}^{\prime}_{Z} &=
&\frac{1}{I^{\prime}} \int d^{2}H
\left<
  \frac{1}{2} \vecbf{E}^{\prime T}(\vecbf{H};t) \cdot
  \matbf{J}_{Z}(\vecbf{H}) \cdot \vecbf{E}^{\prime *}(\vecbf{H};t)
\right> \, = \,
\frac{1}{I^{\prime}}
\int d^{2}H \left<
  \frac{1}{2} \left[\matbf{D} \cdot \vecbf{E}(\vecbf{H};t)\right]^{T}
  \cdot \matbf{J}_{Z}(\vecbf{H}) \cdot
  \left[\matbf{D} \cdot \vecbf{E}(\vecbf{H};t)\right]^{*}
\right> \nonumber \\ &=
&\frac{1}{I^{\prime}}
\int d^{2}H \left<
  \frac{1}{2} \vecbf{E}^{T}(\vecbf{H};t) \cdot
  \left[
    \matbf{D}^{T} \cdot \matbf{J}_{Z}(\vecbf{H}) \cdot \matbf{D}^{*}
  \right]
  \cdot \vecbf{E}^{*}(\vecbf{H};t)
\right> \, = \,
\frac{1}{I^{\prime}}
\int d^{2}H \left<
  \frac{1}{2} \vecbf{E}^{T}(\vecbf{H};t) \cdot
  \matbf{J}^{\prime}_{Z}(\vecbf{H}) \cdot \vecbf{E}^{*}(\vecbf{H};t)
\right> \, ,
\end{eqnarray}
where $\matbf{J}^{\prime}_{Z}$ is the operator that includes mixing effects
from the imperfect instrument,
\begin{eqnarray} \label{Eq:IprimeImperfect}
I^{\prime} &=
&\int d^{2}H
\left<
  \frac{1}{2} \vecbf{E}^{\prime T}(\vecbf{H};t) \cdot
  \vecbf{E}^{\prime *}(\vecbf{H};t)
\right> \, = \,
\int d^{2}H
\left<
  \frac{1}{2} \left[\matbf{D} \cdot \vecbf{E}(\vecbf{H};t)\right]^{T} \cdot
  \left[\matbf{D} \cdot \vecbf{E}(\vecbf{H};t)\right]^{*}
\right> \nonumber \\ &=
&\int d^{2}H
\left<
  \frac{1}{2} \vecbf{E}^{T}(\vecbf{H};t) \cdot
  \left[ \matbf{D}^{T} \cdot \matbf{D}^{*} \right]
  \cdot \vecbf{E}^{*}(\vecbf{H};t)
\right> \, = \,
\int d^{2}H
\left<
  \frac{1}{2} \vecbf{E}^{T}(\vecbf{H};t) \cdot \matbf{\mathcal{D}}_{0} \cdot
  \vecbf{E}^{*}(\vecbf{H};t)
\right> \, ,
\end{eqnarray}
is the integrated intensity through the imperfect instrument, and
\begin{equation} \label{Eq:D0D}
\matbf{\mathcal{D}}_{0} \, = \,
\matbf{D}^{T} \, \cdot \, \matbf{D}^{*} \, = \,
\left[
  \begin{array}{cc}
    \left|D^{A,A}\right|^{2} + \left|D^{B,A}\right|^{2} &
    D^{A,A} D^{A,B *} + D^{B,A} D^{B,B *} \\
    D^{A,A *} D^{A,B} + D^{B,A *} D^{B,B} &
    \left|D^{A,B}\right|^{2} + \left|D^{B,B}\right|^{2}
  \end{array}
\right]
\end{equation}
is a 2x2 matrix.
\end{mathletters}
Because of the linearity of the POAM and PSAM operators, the mixed PTAM operator
becomes
\begin{mathletters}
\begin{equation} \label{Eq:JopExpand}
\matbf{J}^{\prime}_{Z}(\vecbf{H}) \, = \,
  \matbf{D}^{T} \cdot \matbf{J}_{Z}(\vecbf{H}) \cdot \matbf{D}^{*}
    \, = \,
  \matbf{D}^{T} \cdot \matbf{L}_{Z}(\vecbf{H}) \cdot \matbf{D}^{*}
    \, + \,
  \matbf{D}^{T} \cdot \matbf{S}_{Z}(\vecbf{H}) \cdot \matbf{D}^{*}
    \, = \,
  \matbf{L}^{\prime}_{Z}(\vecbf{H}) \, + \, \matbf{S}^{\prime}_{Z}(\vecbf{H})
    \, ,
\end{equation}
where
\begin{equation} \label{Eq:LopExpand}
\matbf{L}^{\prime}_{Z}(\vecbf{H}) \, = \,
\left[
  \matbf{D}^{T} \cdot \matbf{\sigma}_{0} \cdot \matbf{D}^{*}
\right] \,
  j \hbar \frac{\partial}{\partial \chi} \, = \,
\left[\matbf{D}^{T} \cdot \matbf{D}^{*}\right] \,
  j \hbar \frac{\partial}{\partial \chi} \, = \,
\matbf{\mathcal{D}}_{0} \, j \hbar \frac{\partial}{\partial \chi}
\end{equation}
is the POAM operator including the imperfect instrument,
\begin{equation} \label{Eq:SopExpand}
\matbf{S}^{\prime}_{Z}(\vecbf{H}) \, = \,
\left[\matbf{D}^{T} \cdot \matbf{\sigma}_{3} \cdot \matbf{D}^{*}\right]
  \, \hbar \, = \,
\matbf{\mathcal{D}}_{3} \, \hbar
\end{equation}
is the PSAM operator including the imperfect instrument,
and
\begin{equation} \label{Eq:D3D}
\matbf{\mathcal{D}}_{3} \, = \,
 \left[
  \begin{array}{cc}
    \left|D^{A,A}\right|^{2} - \left|D^{B,A}\right|^{2} &
    D^{A,A} D^{A,B *} - D^{B,A} D^{B,B *} \\
    D^{A,A *} D^{A,B} - D^{B,A *} D^{B,B} &
    \left|D^{A,B}\right|^{2} - \left|D^{B,B}\right|^{2}
  \end{array}
\right]
\end{equation}
is another 2x2 matrix.  When $\matbf{D}$ $=$ $\eta$ $\matbf{1}$:
$\matbf{\mathcal{D}}_{0}$ $=$ $\left|\eta\right|^{2}$ $\matbf{\sigma}_{0}$ $=$
$\left|\eta\right|^{2}$ $\matbf{1}$, $\matbf{\mathcal{D}}_{3}$ $=$
$\left|\eta\right|^{2}$ $\matbf{\sigma}_{3}$,
$\matbf{J}^{\prime}_{Z}(\vecbf{H})$ $=$ $\left|\eta\right|^{2}$
$\matbf{J}_{Z}(\vecbf{H})$, $\matbf{L}^{\prime}_{Z}(\vecbf{H})$ $=$
$\left|\eta\right|^{2}$ $\matbf{L}_{Z}(\vecbf{H})$,
$\matbf{S}^{\prime}_{Z}(\vecbf{H})$ $=$ $\left|\eta\right|^{2}$
$\matbf{S}_{Z}(\vecbf{H})$, $\hat{J}^{\prime}_{Z}$ $=$ $\hat{J}_{Z}$,
$\hat{L}^{\prime}_{Z}$ $=$ $\hat{L}_{Z}$, and $\hat{S}^{\prime}_{Z}$ $=$
$\hat{S}_{Z}$.  The $\eta$ factor does not modify the expectation values because
they are normalized quantities.
\end{mathletters}

\section{Simple Example} \label{Sec:SimpleExample}

In Section \ref{Sec:Introduction}, I show that the most general unfactored PTAM
electric field has PSAM states with different POAM expansions.  In this section,
I demonstrate how the measured POAM expectation value can be affected by source
and instrumental PSAM using the simplest unfactored PTAM electric field
\begin{equation} \label{Eq:EFieldSimple}
\vecbf{E}(\vecbf{H};t) \, = \,
  \left[
    \begin{array}{c}
      E_{A}(\vecbf{H};t) \\
      E_{B}(\vecbf{H};t)
    \end{array}
  \right] \, = \,
  \left[
    \begin{array}{c}
      E_{A,m}(H;t) \, e^{j m \chi} \\
      E_{B,n}(H;t) \, e^{j n \chi}
    \end{array}
  \right] \, .
\end{equation}
The PTAM expectation value ultimately depends on the behavior of three
intensity-based quantities,
\begin{mathletters}
\begin{eqnarray} \label{Eq:I_AA}
I_{A,A} &=
&\int d^{2}H \, I_{A,A}(\vecbf{H}) \, = \,
  \int d^{2}H \, \left<
    \frac{1}{2} \left|E_{A}(\vecbf{H};t)\right|^{2}
  \right> \nonumber \\ &=
&2 \pi \int dH \, H \, I_{A,A,m,m}(H) \, = \,
  2 \pi \int dH \, H \, \left<
    \frac{1}{2} \left|E_{A,m}(H;t)\right|^{2}
  \right> \, ,
\end{eqnarray}
\begin{eqnarray} \label{Eq:I_BB}
I_{B,B} &=
&\int d^{2}H \, I_{B,B}(\vecbf{H}) \, = \,
  \int d^{2}H \, \left<
    \frac{1}{2} \left|E_{B}(\vecbf{H};t)\right|^{2}
  \right> \nonumber \\ &=
&2 \pi \int dH \, H \, I_{B,B,n,n}(H) \, = \,
  2 \pi \int dH \, H \, \left<
    \frac{1}{2} \left|E_{B,n}(H;t)\right|^{2}
  \right> \, ,
\end{eqnarray}
and
\begin{eqnarray} \label{Eq:I_AB}
I_{A,B} \, \delta_{m,n} &=
&\int d^{2}H \, I_{A,B}(\vecbf{H}) \, = \,
  \int d^{2}H \, \left<
    \frac{1}{2} E_{A}(\vecbf{H};t) \, E^{*}_{B}(\vecbf{H};t)
  \right> \nonumber \\ &=
&2 \pi \int dH \, H \, I_{A,B,m,n}(H) \, \delta_{m,n} =
  2 \pi \int dH \, H \, \left<
    \frac{1}{2} E_{A,m}(H;t) \, E^{*}_{B,n}(H;t)
  \right> \, \delta_{m,n} \, ,
\end{eqnarray}
where $\delta_{m,n}$ is the Kronecker delta function.  They can be rearranged
to become more familiar quantities, namely the Stokes parameters $I$ $=$
$I_{A,A}$ $+$ $I_{B,B}$, $Q$ $=$ $2$ $\mathbf{Re} \left\{I_{A,B}\right\}$
$\delta_{m,n}$, $U$ $=$ $2$ $\mathbf{Im} \left\{I_{A,B}\right\}$ $\delta_{m,n}$,
and $V$ $=$ $I_{A,A}$ $-$ $I_{B,B}$.  This electric field contains linear
polarization only when the PSAM states have the same POAM state, or $m$ $=$ $n$.
\end{mathletters}

Using the definitions in Sections \ref{Sec:E_PTAM} and
\ref{Sec:ExpectationValues}, the PTAM expectation value becomes
\begin{equation} \label{Eq:JExpSimple}
\hat{J}_{Z} \, = \,
  \hat{L}_{Z} \, + \, \hat{S}_{Z} \, = \,
  \left[
    \left\{ m \frac{I_{A,A}}{I} \, + \, n \frac{I_{B,B}}{I} \right\} \, + \,
    \left\{ \frac{I_{A,A}}{I} - \frac{I_{B,B}}{I} \right\}
  \right] \, \hbar \, = \,
  \left[
    \left\{ m p_{A,A} + n p_{B,B} \right\} \, + \,
    \left\{ p_{A,A} - p_{B,B} \right\}
  \right] \, \hbar \, = \,
  \left[
    M_{eff} \, + \, v
  \right] \, \hbar \, .
\end{equation}
Note that $p_{A,A}$ and $p_{B,B}$ are part of both the POAM and PSAM expectation
values.  For a purely unpolarized and/or linearly polarized source $p_{A,A}$ $=$
$p_{B,B}$ $=$ $\half$ and $\hat{J}_{Z}$ $=$ $\half \left( m + n \right) \hbar$.
Also, $\hat{J}_{Z}$ $=$ $\left( m + 1 \right) \hbar$ for a purely right-handed
circularly polarized source ($p_{A,A}$ $=$ $1$ and $p_{B,B}$ $=$ $0$) and
$\hat{J}_{Z}$ $=$ $\left( n - 1 \right) \hbar$ for a purely left-handed
circularly polarized source ($p_{A,A}$ $=$ $0$ and $p_{B,B}$ $=$ $1$).  If $m$
$=$ $n$ (factored electric field), those PTAM expectation values become
$\hat{J}_{Z}$ $=$ $m \hbar$ (PTAM expectation value is independent of PSAM
expectation value), $\left( m + 1 \right) \hbar$ (PTAM expectation value is POAM
expectation value plus RCP PSAM expectation value), and $\left( m - 1 \right)
\hbar$ (PTAM expectation value is POAM expectation value minus LCP PSAM
expectation value), respectively.  

An instrument with non-zero instrumental PSAM modifies the result of Equation
\ref{Eq:JExpSimple}.  Using a Jones matrix of the form in Equation
\ref{Eq:JonesMatrixSimple}, the electric field becomes
\begin{equation} \label{Eq:EFieldSimple_D}
\vecbf{E}^{\prime}(\vecbf{H};t) \, = \,
  \left[
    \begin{array}{c}
      E^{\prime}_{A}(\vecbf{H};t) \\
      E^{\prime}_{B}(\vecbf{H};t)
    \end{array}
  \right] \, = \,
  \matbf{D} \, \cdot \, \vecbf{E}(\vecbf{H};t) \, = \,
  \left[
    \begin{array}{c}
      D^{A,A} \, E_{A}(\vecbf{H};t) \, + \, D^{A,B} \, E_{B}(\vecbf{H};t) \\
      D^{B,A} \, E_{A}(\vecbf{H};t) \, + \, D^{B,B} \, E_{B}(\vecbf{H};t) \\
    \end{array}
  \right] \, .
\end{equation}
When $\matbf{D}$ $\neq$ $\eta$ $\matbf{1}$ ($D^{A,B}$ $=$ $D^{B,A}$ $=$ $0$),
the instrument mixes the electric field components.  The integrated intensity of
this electric field (cf. Equation \ref{Eq:IprimeImperfect}) can be rewritten in
terms of the source Stokes parameters
\begin{mathletters}
\begin{eqnarray} \label{Eq:IprimeSimple}
I^{\prime} &=
&\int d^{2}H \, I^{\prime}(\vecbf{H}) \, = \,
  \int d^{2}H \, \left<
    \frac{1}{2} \vecbf{E}^{T}(\vecbf{H};t)
    \cdot \matbf{\mathcal{D}}_{0} \cdot
    \vecbf{E}^{*}(\vecbf{H};t)
  \right> \nonumber \\ &=
&\left[ \left|D^{A,A}\right|^{2} + \left|D^{B,A}\right|^{2} \right] \, I_{A,A}
  \, + \,
  \left[ \left|D^{A,B}\right|^{2} + \left|D^{B,B}\right|^{2} \right] \, I_{B,B}
  \, + \,
  2 \mathbf{Re} \, \left\{
    \left[ D^{A,A} D^{A,B *} + D^{B,A} D^{B,B *} \right] \, I_{A,B}
  \right\} \, \delta_{m,n} \nonumber \\ &=
&M^{I,I} \, I \, + \, M^{I,Q} \, \delta_{m,n} \, Q \, + \,
  M^{I,U} \, \delta_{m,n} \, U \, + \, M^{I,V} \, V \, = \,
\left[ M^{I,I} \, + \, M^{I,Q} \, \delta_{m,n} \, q \, + \,
  M^{I,U} \, \delta_{m,n} \, u \, + \, M^{I,V} \, v \right] \, I \, = \,
\mathcal{M} \, I \, ,
\end{eqnarray}
where
\begin{equation} \label{eq:MII}
M^{I,I} \, = \,
  \frac{1}{2} \left[
    \left|D^{A,A}\right|^{2} \, + \, \left|D^{B,A}\right|^{2} \, + \,
    \left|D^{A,B}\right|^{2} \, + \, \left|D^{B,B}\right|^{2}
  \right] \, ,
\end{equation}
\begin{equation} \label{eq:MIQ}
M^{I,Q} \, = \,
  \mathbf{Re} \left[ D^{A,A} D^{A,B *} + D^{B,A} D^{B,B *} \right] \, ,
\end{equation}
\begin{equation} \label{Eq:MIU}
M^{I,U} \, = \,
  - \mathbf{Im} \left[ D^{A,A} D^{A,B *} + D^{B,A} D^{B,B *} \right]
\end{equation}
and
\begin{equation} \label{Eq:MIV}
M^{I,V} \, = \,
  \frac{1}{2} \left[
    \left|D^{A,A}\right|^{2} \, + \, \left|D^{B,A}\right|^{2} \, - \,
    \left|D^{A,B}\right|^{2} \, - \, \left|D^{B,B}\right|^{2}
  \right]
\end{equation}
are instrument-dependent coefficients, and $q$ $=$ $Q/I$, $u$ $=$ $U/I$, and $v$
$=$ $V/I$ are the normalized source Stokes parameters.  These equations are
interesting, because 1) source circular polarization contributes to
$I^{\prime}$; and 2) if $m$ $=$ $n$ (unfactored electric field) the linear PSAM
contributes to $I^{\prime}$.  Note that the system gain $\mathcal{M}$
$\rightarrow$ $\mathcal{M}(q,u,v)$, i.e., it depends on both the instrumental
and source PSAM.
\end{mathletters}

Using Equations \ref{Eq:EFieldSimple_D} and \ref{Eq:IprimeSimple} and the
mathematics of Section \ref{SSec:ImperfectInstrument}, the POAM and PSAM
expectation values including instrumental PSAM are
\begin{mathletters}
\begin{eqnarray} \label{Eq:LprimeSimple}
\hat{L}^{\prime}_{Z} &=
&\frac{M^{I,I} + M^{I,V}}{\mathcal{M}} p_{A,A} \, m \hbar +
  \frac{M^{I,I} - M^{I,V}}{\mathcal{M}} p_{B,B} \, n \hbar +
  \frac{M^{I,Q} - j M^{I,U}}{\mathcal{M}} p_{A,B} \, n \hbar \, \delta_{m,n} +
  \frac{M^{I,Q} + j M^{I,U}}{\mathcal{M}} p^{*}_{A,B} \, m \hbar \, \delta_{m,n}
  \nonumber \\ &=
&\hat{L}_{Z} \, + \, \left[
  \frac{M^{I,I} + M^{I,V} - \mathcal{M}}{\mathcal{M}} p_{A,A} \, m \hbar +
  \frac{M^{I,I} - M^{I,V} - \mathcal{M}}{\mathcal{M}} p_{B,B} \, n \hbar +
  \frac{M^{I,Q} - j M^{I,U}}{\mathcal{M}} p_{A,B} \, n \hbar \, \delta_{m,n} +
  \frac{M^{I,Q} + j M^{I,U}}{\mathcal{M}} p^{*}_{A,B} \, m \hbar \, \delta_{m,n}
\right] \nonumber \\ &=
&\hat{L}_{Z} \, + \, \Delta \hat{L}_{Z}
\end{eqnarray}
and
\begin{eqnarray} \label{Eq:SprimeSimple}
\hat{S}^{\prime}_{Z} &=
&\frac{M^{V,I}}{\mathcal{M}} \, \hbar \, + \,
  \frac{M^{V,Q}}{\mathcal{M}} \, q \hbar \, \delta_{m,n} \, + \,
  \frac{M^{V,U}}{\mathcal{M}} \, u \hbar \, \delta_{m,n} \, + \,
  \frac{M^{V,V}}{\mathcal{M}} \, v \hbar \nonumber \\ &=
&\hat{S}_{Z} \, + \, \left[ \frac{M^{V,I}}{\mathcal{M}} \, \hbar \, + \,
  \frac{M^{V,Q}}{\mathcal{M}} \, q \hbar \, \delta_{m,n} \, + \,
  \frac{M^{V,U}}{\mathcal{M}} \, u \hbar \, \delta_{m,n} \, + \,
  \frac{M^{V,V} - \mathcal{M}}{\mathcal{M}} \, v \hbar \right] \nonumber \\ &=
&\hat{S}_{Z} \, + \, \Delta \hat{S}_{Z} \, ,
\end{eqnarray}
where
\begin{equation} \label{Eq:MVI}
M^{V,I} \, = \,
  \frac{1}{2} \left[
    \left|D^{A,A}\right|^{2} \, - \, \left|D^{B,A}\right|^{2} \, + \,
    \left|D^{A,B}\right|^{2} \, - \, \left|D^{B,B}\right|^{2}
  \right] \, ,
\end{equation}
\begin{equation} \label{eq:MVQ}
M^{V,Q} \, = \,
  \mathbf{Re} \left[ D^{A,A} D^{A,B *} - D^{B,A} D^{B,B *} \right] \, ,
\end{equation}
\begin{equation} \label{Eq:MVU}
M^{V,U} \, = \,
  - \mathbf{Im} \left[ D^{A,A} D^{A,B *} - D^{B,A} D^{B,B *} \right] \, ,
\end{equation}
and
\begin{equation} \label{Eq:MVV}
M^{V,V} \, = \,
  \frac{1}{2} \left[
    \left|D^{A,A}\right|^{2} \, - \, \left|D^{B,A}\right|^{2} \, - \,
    \left|D^{A,B}\right|^{2} \, + \, \left|D^{B,B}\right|^{2}
  \right]
\end{equation}
are other instrument-dependent coefficients, and $p_{A,B}$ $=$ $I_{A,B}$ $/$ $I$
$=$ $\half \left(q + j u\right)$ is the ``transitional probability.''  Equation
\ref{Eq:SprimeSimple} shows that instrumental PSAM changes the PSAM expectation
value.  \citet{Elias1} showed that instrumental POAM changes the POAM
expectation value.  These results are not unexpected and not particularly
exciting.  On the other hand, changes in the POAM expectation value due to
instrumental and source PSAM deserve further analysis.  I call this effect
``P\it{\underline{S}}\rm{AM-}\it{\underline{M}}\rm{odified P}\it{\underline{O}}\rm{AM }\it{\underline{M}}\rm{easurements,'' or SMOM}.
\end{mathletters}

After analyzing the complete set of use cases (unpolarized, linearly polarized,
circularly polarized, elliptically polarized source and instrument polarization;
$m$ $\neq$ $n$ or $m$ $=$ $n$), I found that
\begin{equation} \label{Eq:Mixing}
\Delta \hat{L}_{Z} \, = \,
  \frac{1}{2} \frac{m^{I,V}}{1 + m^{I,V} \, v}
  \left( 1 - v^{2} \right) \, \left( m - n \right) \hbar \, ,
\end{equation}
where $m^{I,V}$ $=$ $M^{I,V}$ $/$ $M^{I,I}$ is the normalized circular PSAM
gain.  This equation completely describes the conditions required for SMOM in
this simplified example.

SMOM is possible only when $m$ $\neq$ $n$, i.e., for unfactored PTAM electric
fields.  If the PTAM electric field is factored ($m$ $=$ $n$),
\begin{equation} \label{Eq:ENoMix}
\vecbf{E}(\vecbf{H};t) \, = \,
  \left[
    \begin{array}{c}
      E_{A}(H;t) \\
      E_{B}(H;t)
    \end{array}
  \right] e^{j m \chi} \, ,
\end{equation}
it does not lead to SMOM.  SMOM only occurs in the presence of instrumental
circular PSAM, or $m^{I,V}$ $\neq$ $0$, because only it can mix the different
PSAM states leading to modified POAM expectation values.  No PSAM or
partial/full linear PSAM correspond to $v$ $=$ $0$, which leads to
the maximum $\Delta \hat{L}_{Z}$ for a given instrumental circular PSAM.  The
$0$ $<$ $v$ $<$ $1$ cases correspond to circular source PSAM plus a combination
of unpolarized and/or linear PSAM.  Increasing $v$ decreases the $\Delta
\hat{L}_{Z}$.  The $v$ $=$ $+1$ and $v$ $=$ $-1$ cases correspond to
\begin{mathletters}
\begin{equation} \label{Eq:EAllvPlus1}
\vecbf{E}(\vecbf{H};t) \, = \,
  \left[
    \begin{array}{c}
      E_{A}(H;t) \, e^{j m \chi} \\
      0
    \end{array}
  \right] \, = \,
  \left[
    \begin{array}{c}
      E_{A}(H;t) \\
      0
    \end{array}
  \right] \, e^{j m \chi}
\end{equation}
and
\begin{equation} \label{Eq:EAllvMinus1}
\vecbf{E}(\vecbf{H};t) \, = \,
  \left[
    \begin{array}{c}
      0 \\
      E_{B}(H;t) \, e^{j m \chi}
    \end{array}
  \right] \, = \,
  \left[
    \begin{array}{c}
      0 \\
      E_{B}(H;t)
    \end{array}
  \right] \, e^{j m \chi} \, ,
\end{equation}
respectively.  Because only one PSAM component is non-zero, no SMOM is possible.
These electric fields are similar to Equation \ref{Eq:ENoMix} since the POAM
exponential can also be factored outside the PSAM vector.
\end{mathletters}

\section{Conclusions} \label{Sec:Conclusions}

I present the most general ``unfactored'' PTAM electric field form, where each
PSAM component has its own POAM expansion.  It is slightly more general than the
more commonly invoked ``factored'' PTAM electric field form where the PSAM and
POAM components are separable.  I then combine the POAM and PSAM calculi to
obtain the PTAM calculi.  Apart from the vectors, matrices, dot products, and
direct products, the PTAM and POAM calculi appear superficially identical.  I
derive the PTAM operator and expectation value in terms of POAM/PSAM operators
and expectation values for systems with and without instrumental PSAM.  Last, I
prove using a simple example that POAM measurements of sources with unfactored
PTAM electric fields passing through instrumental circular PSAM yield systematic
POAM measurement errors.

\begin{table}[!ht]

\begin{center}
\begin{minipage}[!ht]{11.0cm}
\caption{The POAM expansions of $\vecbf{E}(\vecbf{\Omega}^{\prime};\vecbf{a},t)$
in terms of the POAM expansions of
$\matbf{D}(\vecbf{\Omega}^{\prime},\vecbf{\Omega};\vecbf{a})$ (Table
\ref{Tab:ExpD}) and $\vecbf{E}(\vecbf{\Omega};t)$.  The vector $\vecbf{a}$ is
a generic representation of optional configuration parameters.}
\begin{tabular}{|l|l|}
\hline \hline
POAM expansion type & Expression \\
\hline \hline
Input & $\vecbf{E}(\vecbf{\Omega}^{\prime};\vecbf{a},t)$ $=$
$\sum_{m=-\infty}^{\infty}
\vecbf{\mathfrak{E}}_{m}(\vecbf{\Omega}^{\prime};\vecbf{a},t)$ \\
& where $\vecbf{\mathfrak{E}}_{m}(\vecbf{\Omega}^{\prime};\vecbf{a},t)$ $=$
  $2\pi \int_{0}^{\infty} d\rho \, \rho \,
  \matbf{D}^{-m}(\vecbf{\Omega}^{\prime},\rho;\vecbf{a}) \, \cdot \,
  \vecbf{E}_{m}(\rho;t)$ \\
\hline \hline
Output & $\vecbf{E}(\vecbf{\Omega}^{\prime};\vecbf{a},t)$ $=$
  $\sum_{p=-\infty}^{\infty}
  \vecbf{E}_{p}(\rho^{\prime};\vecbf{a},t) \, e^{j p \phi^{\prime}}$ \\
& where $\vecbf{E}_{p}(\rho^{\prime};\vecbf{a},t)$ $=$ $\int d^{2}\Omega \,
  \matbf{D}_{p}(\rho^{\prime},\vecbf{\Omega};\vecbf{a}) \, \cdot \,
  \vecbf{E}(\vecbf{\Omega};t)$ \\
\hline \hline
Input/Output & $\vecbf{E}(\vecbf{\Omega}^{\prime};\vecbf{a},t)$ $=$
  $\sum_{p=-\infty}^{\infty} \vecbf{E}_{p}(\rho^{\prime};\vecbf{a},t) \,
  e^{j p \phi^{\prime}}$ \\
& where $\vecbf{E}_{p}(\rho^{\prime};\vecbf{a},t)$ $=$
  $\sum_{m=-\infty}^{\infty} 2\pi \int_{0}^{\infty} d\rho \, \rho \,
  \matbf{D}^{-m}_{p}(\rho^{\prime},\rho;\vecbf{a}) \, \cdot \,
  \vecbf{E}_{m}(\rho;t)$ \\
\hline \hline
\end{tabular}
\label{Tab:ExpE}
\end{minipage}
\end{center}

\begin{center}
\begin{minipage}[!ht]{10.25cm}
\caption{The POAM expansions of the
$\matbf{D}(\vecbf{\Omega}^{\prime},\vecbf{\Omega};\vecbf{a})$.}
\begin{tabular}{|l|l|}
\hline \hline
POAM expansion & Expression \\
\hline \hline
Input sensitivity: & \\
Integral form (forward) &
  $\matbf{D}^{-m}(\vecbf{\Omega}^{\prime},\rho;\vecbf{a})$ $=$
  $\frac{1}{2\pi} \int_{0}^{2\pi} d\phi \, e^{j m \phi} \,
  \matbf{D}(\vecbf{\Omega}^{\prime},\vecbf{\Omega};\vecbf{a})$ \\
Sum form (reverse) &
  $\matbf{D}(\vecbf{\Omega}^{\prime},\vecbf{\Omega};\vecbf{a})$ $=$
  $\sum_{m=-\infty}^{\infty}
  \matbf{D}^{-m}(\vecbf{\Omega}^{\prime},\rho;\vecbf{a}) \, e^{-j m \phi}$ \\
\hline \hline
Output sensitivity: & \\
Integral form (forward) &
  $\matbf{D}_{p}(\rho^{\prime},\vecbf{\Omega};\vecbf{a})$ $=$
  $\frac{1}{2\pi} \int_{0}^{2\pi} d\phi^{\prime} \, e^{-j p \phi^{\prime}} \,
  \matbf{D}(\vecbf{\Omega}^{\prime},\vecbf{\Omega};\vecbf{a})$ \\
Sum form (reverse) &
  $\matbf{D}(\vecbf{\Omega}^{\prime},\vecbf{\Omega};\vecbf{a})$ $=$
  $\sum_{p=-\infty}^{\infty}
  \matbf{D}_{p}(\rho^{\prime},\vecbf{\Omega};\vecbf{a}) \,
  e^{j p \phi^{\prime}}$ \\
\hline \hline
Input/Output gain: & \\
Integral form (forward) & $\matbf{D}^{-m}_{p}(\rho^{\prime},\rho;\vecbf{a})$ $=$
  $\frac{1}{2\pi} \int_{0}^{2\pi} d\phi \, e^{j m \phi} \, \frac{1}{2\pi}
  \int_{0}^{2\pi} \, d\phi^{\prime} \, e^{-j p \phi^{\prime}} \,
  \matbf{D}(\vecbf{\Omega}^{\prime},\vecbf{\Omega};\vecbf{a})$ \\
Sum form (reverse) &
  $\matbf{D}(\vecbf{\Omega}^{\prime},\vecbf{\Omega};\vecbf{a})$ $=$
  $\sum_{p=-\infty}^{\infty} \sum_{m=-\infty}^{\infty}
  \matbf{D}^{-m}_{p}(\rho^{\prime},\rho;\vecbf{a}) \, e^{-j m \phi} \,
  e^{j p \phi^{\prime}}$ \\
\hline \hline
\end{tabular}
\label{Tab:ExpD}
\end{minipage}
\end{center}

\end{table}

\begin{table}[!ht]

\begin{center}
\begin{minipage}[!ht]{12.0cm}
\caption{The POAM expansions of $\vecbf{S}(\vecbf{\Omega}^{\prime};\vecbf{a})$,
for a spatially incoherent source, in terms of the POAM expansions of
$\matbf{P}(\vecbf{\Omega}^{\prime},\vecbf{\Omega};\vecbf{a})$ (Table
\ref{Tab:ExpP}) and $\vecbf{S}(\vecbf{\Omega})$.  The vector $\vecbf{a}$ is
a generic representation of optional configuration parameters.}
\begin{tabular}{|l|l|}
\hline \hline
POAM expansion type & Expression \\
\hline \hline
Input &
  $\vecbf{S}(\vecbf{\Omega}^{\prime};\vecbf{a})$ $=$
  $\sum_{m=-\infty}^{\infty} \sum_{n=-\infty}^{\infty}
  \vecbf{\mathfrak{S}}_{m,n}(\vecbf{\Omega}^{\prime};\vecbf{a})$ \\
(correlated) & where
  $\vecbf{\mathfrak{S}}_{m,n}(\vecbf{\Omega}^{\prime};\vecbf{a})$ $=$
  $2\pi \int_{0}^{\infty} d\rho \, \rho \,
  \matbf{\mathcal{P}}^{-m+n}(\vecbf{\Omega}^{\prime},\rho;\vecbf{a}) \, \cdot \,
  \vecbf{S}_{m,n}(\rho)$ \\
\hline
Input &
  $\vecbf{S}(\vecbf{\Omega}^{\prime};\vecbf{a})$ $=$ $\sum_{m=-\infty}^{\infty}
  \vecbf{\mathfrak{S}}_{m}(\vecbf{\Omega}^{\prime};\vecbf{a})$ \\
(rancored) & where $\vecbf{\mathfrak{S}}_{m}(\vecbf{\Omega}^{\prime};\vecbf{a})$
  $=$ $2\pi \int_{0}^{\infty} d\rho \, \rho \,
  \matbf{\mathcal{P}}^{-m}(\vecbf{\Omega}^{\prime},\rho;\vecbf{a}) \, \cdot \,
  \vecbf{\mathcal{S}}_{m}(\rho)$ \\
\hline \hline
Output &
  $\vecbf{S}(\vecbf{\Omega}^{\prime};\vecbf{a})$ $=$ $\sum_{p=-\infty}^{\infty}
  \sum_{q=-\infty}^{\infty} \vecbf{S}_{p,q}(\rho^{\prime};\vecbf{a}) \,
  e^{j (p-q) \phi^{\prime}}$ \\
(correlated/unexpanded) & where $\vecbf{S}_{p,q}(\rho^{\prime};\vecbf{a})$ $=$
$\int d^{2}\Omega \, \matbf{P}_{p,q}(\rho^{\prime},\vecbf{\Omega};\vecbf{a})
  \, \cdot \, \vecbf{S}(\vecbf{\Omega})$ \\
\hline
Output &
  $\vecbf{S}(\vecbf{\Omega}^{\prime};\vecbf{a})$ $=$ $\sum_{p=-\infty}^{\infty}
  \vecbf{\mathcal{S}}_{p}(\rho^{\prime};\vecbf{a}) \, e^{j p \phi^{\prime}}$ \\
(rancored/unexpanded) & where $\vecbf{\mathcal{S}}_{p}(\rho^{\prime};\vecbf{a})$
$=$ $\int d^{2}\Omega \,
  \matbf{\mathcal{P}}_{p}(\rho^{\prime},\vecbf{\Omega};\vecbf{a}) \, \cdot \,
  \vecbf{S}(\vecbf{\Omega})$ \\
\hline \hline
Input/Output &
  $\vecbf{S}(\vecbf{\Omega}^{\prime};\vecbf{a})$ $=$
  $\sum_{p=-\infty}^{\infty} \sum_{q=-\infty}^{\infty}
  \vecbf{S}_{p,q}(\rho^{\prime};\vecbf{a}) \, e^{j (p-q) \phi^{\prime}}$ \\
(correlated/correlated) & where $\vecbf{S}_{p,q}(\rho^{\prime};\vecbf{a})$ $=$
  $\sum_{m=-\infty}^{\infty} \sum_{n=-\infty}^{\infty} 2\pi \int_{0}^{\infty}
  d\rho \, \rho \,
  \matbf{\mathcal{P}}^{-m+n}_{p,q}(\rho^{\prime},\rho;\vecbf{a}) \, \cdot \,
  \vecbf{S}_{m,n}(\rho)$\\
\hline
Input/Output &
  $\vecbf{S}(\vecbf{\Omega}^{\prime};\vecbf{a})$ $=$ $\sum_{p=-\infty}^{\infty}
  \sum_{q=-\infty}^{\infty} \vecbf{S}_{p,q}(\rho^{\prime};\vecbf{a}) \,
  e^{j (p-q) \phi^{\prime}}$ \\
(correlated/rancored) & where $\vecbf{S}_{p,q}(\rho^{\prime};\vecbf{a})$ $=$
  $\sum_{m=-\infty}^{\infty} 2\pi \int_{0}^{\infty} d\rho \, \rho \,
  \matbf{\mathcal{P}}^{-m}_{p,q}(\rho^{\prime},\rho;\vecbf{a}) \, \cdot \,
  \vecbf{\mathcal{S}}_{m}(\rho)$ \\
\hline
Input/Output &
  $\vecbf{S}(\vecbf{\Omega}^{\prime};\vecbf{a})$ $=$ $\sum_{p=-\infty}^{\infty}
  \vecbf{\mathcal{S}}_{p}(\rho^{\prime};\vecbf{a}) \, e^{j p \phi^{\prime}}$ \\
(rancored/correlated) & where $\vecbf{\mathcal{S}}_{p}(\rho^{\prime};\vecbf{a})$
$=$ $\sum_{m=-\infty}^{\infty} \sum_{n=-\infty}^{\infty} 2\pi \int_{0}^{\infty}
  d\rho \, \rho \, \matbf{\mathcal{P}}^{-m+n}_{p}(\rho^{\prime},\rho;\vecbf{a})
  \, \cdot \, \vecbf{S}_{m,n}(\rho)$ \\
\hline
Input/Output &
  $\vecbf{S}(\vecbf{\Omega}^{\prime};\vecbf{a})$ $=$ $\sum_{p=-\infty}^{\infty}
  \vecbf{\mathcal{S}}_{p}(\rho^{\prime};\vecbf{a}) \, e^{j p \phi^{\prime}}$ \\
(rancored/rancored) & where $\vecbf{\mathcal{S}}_{p}(\rho^{\prime};\vecbf{a})$
$=$ $\sum_{m=-\infty}^{\infty} 2\pi \int_{0}^{\infty} d\rho \, \rho \,
  \matbf{\mathcal{P}}^{-m}_{p}(\rho^{\prime},\rho;\vecbf{a}) \, \cdot \,
  \vecbf{\mathcal{S}}_{m}(\rho)$ \\
\hline \hline
\end{tabular}
\label{Tab:ExpS}
\end{minipage}
\end{center}

\begin{center}
\begin{minipage}[!ht]{16.25cm}
\caption{The POAM expansions of
$\matbf{P}(\vecbf{\Omega}^{\prime},\vecbf{\Omega};\vecbf{a})$.}
\begin{tabular}{|l|l|}
\hline \hline
POAM expansion & Expression \\
\hline \hline
Input sensitivity (separate): & \\
Integral form (forward) &
  $\matbf{P}^{-m,-n}(\vecbf{\Omega}^{\prime},\rho;\vecbf{a})$ $=$
  $\matbf{T} \cdot \left[ \matbf{D}^{-m}(\vecbf{\Omega}^{\prime},\rho;\vecbf{a})
  \otimes \matbf{D}^{-n,*}(\vecbf{\Omega}^{\prime},\rho;\vecbf{a}) \right]
  \cdot \matbf{T}^{-1}$ \\
Sum form (reverse) &
  $\matbf{P}(\vecbf{\Omega}^{\prime},\vecbf{\Omega};\vecbf{a})$ $=$
  $\sum_{m=-\infty}^{\infty} \sum_{n=-\infty}^{\infty}$ 
  $\matbf{P}^{-m,-n}(\vecbf{\Omega}^{\prime},\rho;\vecbf{a})$
  $e^{-j(m-n) \phi}$ \\
\hline
Output sensitivity (separate): & \\
Integral form (forward) &
  $\matbf{P}_{p,q}(\rho^{\prime},\vecbf{\Omega};\vecbf{a})$ $=$
  $\matbf{T} \cdot \left[ \matbf{D}_{p}(\rho^{\prime},\vecbf{\Omega};\vecbf{a})
  \otimes \matbf{D}^{*}_{q}(\rho^{\prime},\vecbf{\Omega};\vecbf{a}) \right]
  \cdot \matbf{T}^{-1}$ \\
Sum form (reverse) &
  $\matbf{P}(\vecbf{\Omega}^{\prime},\vecbf{\Omega};\vecbf{a})$ $=$
  $\sum_{p=-\infty}^{\infty} \sum_{q=-\infty}^{\infty}$
  $\matbf{P}_{p,q}(\rho^{\prime},\vecbf{\Omega};\vecbf{a})$
  $e^{j(p-q) \phi^{\prime}}$ \\
\hline
Input/Output gain (separate): & \\
Integral form (forward) &
  $\matbf{P}^{-m,-n}_{p,q}(\rho^{\prime},\rho;\vecbf{a})$ $=$
  $\matbf{T} \cdot \left[ \matbf{D}^{-m}_{p}(\rho^{\prime},\rho;\vecbf{a})
  \otimes \matbf{D}^{-n,*}_{q}(\rho^{\prime},\rho;\vecbf{a}) \right] \cdot
  \matbf{T}^{-1}$ \\
Sum form (reverse) &
  $\matbf{P}(\vecbf{\Omega}^{\prime},\vecbf{\Omega};\vecbf{a})$ $=$
  $\sum_{p=-\infty}^{\infty} \sum_{q=-\infty}^{\infty}$
  $\sum_{m=-\infty}^{\infty} \sum_{n=-\infty}^{\infty}$ 
  $\matbf{P}_{p,q}^{-m,-n}(\rho^{\prime},\rho;\vecbf{a})$ $e^{-j(m-n)\phi}$
  $e^{j(p-q)\phi^{\prime}}$ \\
\hline \hline
Input sensitivity (combined): & \\
Integral form (forward) &
  $\matbf{\mathcal{P}}^{-m}(\vecbf{\Omega}^{\prime},\rho;\vecbf{a})$ $=$
  $\frac{1}{2\pi} \int_{0}^{2\pi} d\phi \, e^{j m \phi} \,
  \matbf{P}(\vecbf{\Omega}^{\prime},\vecbf{\Omega};\vecbf{a})$ $=$
  $\sum_{k=-\infty}^{\infty}
  \matbf{P}^{-k,-k+m}(\vecbf{\Omega}^{\prime},\rho;\vecbf{a})$ \\
Sum form (reverse) &
  $\matbf{P}(\vecbf{\Omega}^{\prime},\vecbf{\Omega};\vecbf{a})$ $=$
  $\sum_{m=-\infty}^{\infty}
  \matbf{\mathcal{P}}^{-m}(\vecbf{\Omega}^{\prime},\rho;\vecbf{a})$
  $e^{-jm \phi}$ \\
\hline
Output sensitivity (combined): & \\
Integral form (forward) &
  $\matbf{\mathcal{P}}_{p}(\rho^{\prime},\vecbf{\Omega};\vecbf{a})$ $=$
  $\frac{1}{2\pi} \int_{0}^{2\pi} d\phi^{\prime} \, e^{-j p \phi^{\prime}} \,
  \matbf{P}(\vecbf{\Omega}^{\prime},\vecbf{\Omega};\vecbf{a})$ $=$
  $\sum_{l=-\infty}^{\infty}
  \matbf{P}_{l,l-p}(\rho^{\prime},\vecbf{\Omega};\vecbf{a})$ \\
Sum form (reverse) &
  $\matbf{P}(\vecbf{\Omega}^{\prime},\vecbf{\Omega};\vecbf{a})$ $=$
  $\sum_{p=-\infty}^{\infty}
  \matbf{\mathcal{P}}_{p}(\rho^{\prime},\vecbf{\Omega};\vecbf{a})$
  $e^{jp \phi^{\prime}}$ \\
\hline \hline
Input/Output gain (combined \#1): & \\
Integral form (forward) &
  $\matbf{\mathcal{P}}^{-m}_{p}(\rho^{\prime},\rho;\vecbf{a})$ $=$
  $\frac{1}{2\pi} \int_{0}^{2\pi} d\phi \, e^{j m \phi} \, \frac{1}{2\pi}
  \int_{0}^{2\pi} d\phi^{\prime} \, e^{-j p \phi^{\prime}} \,
  \matbf{P}(\vecbf{\Omega}^{\prime},\vecbf{\Omega};\vecbf{a})$ $=$
  $\sum_{k=-\infty}^{\infty} \sum_{l=-\infty}^{\infty}
  \matbf{P}^{-k,-k+m}_{l,l-p}(\rho^{\prime},\rho;\vecbf{a})$ \\
Sum form (reverse) &
  $\matbf{P}(\vecbf{\Omega}^{\prime},\vecbf{\Omega};\vecbf{a})$ $=$
  $\sum_{p=-\infty}^{\infty} \sum_{m=-\infty}^{\infty}
  \matbf{\mathcal{P}}^{-m}_{p}(\rho^{\prime},\rho;\vecbf{a})$ $e^{-jm \phi}$
  $e^{jp \phi^{\prime}}$ \\
\hline
Input/Output gain (combined \#2): & \\
Integral form (forward) &
  $\matbf{\mathcal{P}}^{-m}_{p,q}(\rho^{\prime},\rho;\vecbf{a})$ $=$
  $\frac{1}{2\pi} \int_{0}^{2\pi} d\phi \, e^{j m \phi} \,
  \matbf{P}_{p,q}(\rho^{\prime},\vecbf{\Omega};\vecbf{a})$ $=$
  $\sum_{k=-\infty}^{\infty}
  \matbf{P}^{-k,-k+m}_{p,q}(\rho^{\prime},\rho;\vecbf{a})$ \\
Sum form (reverse) &
  $\matbf{P}(\vecbf{\Omega}^{\prime},\vecbf{\Omega};\vecbf{a})$ $=$
  $\sum_{p=-\infty}^{\infty} \sum_{q=-\infty}^{\infty}$
  $\sum_{m=-\infty}^{\infty}$
  $\matbf{\mathcal{P}}_{p,q}^{-m}(\rho^{\prime},\rho;\vecbf{a})$
  $e^{-jm \phi}$ $e^{j(p-q)\phi^{\prime}}$ \\
\hline
Input/Output gain (combined \#3): & \\
Integral form (forward) & $\matbf{\mathcal{P}}^{-m,-n}_{p}(\rho^{\prime},\rho)$
  $=$ $\frac{1}{2\pi} \int_{0}^{2\pi} d\phi^{\prime} \, e^{-j p \phi^{\prime}}
  \, \matbf{P}^{-m,-n}(\vecbf{\Omega}^{\prime},\rho;\vecbf{a})$ $=$
  $\sum_{l=-\infty}^{\infty}
  \matbf{P}^{-m,-n}_{l,l-p}(\rho^{\prime},\rho;\vecbf{a})$ \\
Sum form (reverse) &
  $\matbf{P}(\vecbf{\Omega}^{\prime},\vecbf{\Omega};\vecbf{a})$ $=$
  $\sum_{p=-\infty}^{\infty} \sum_{m=-\infty}^{\infty}$
  $\sum_{n=-\infty}^{\infty}$
  $\matbf{\mathcal{P}}_{p}^{-m,-n}(\rho^{\prime},\rho;\vecbf{a})$
  $e^{-j(m-n) \phi}$ $e^{jp \phi^{\prime}}$ \\
\hline \hline
\end{tabular}
\label{Tab:ExpP}
\end{minipage}
\end{center}

\end{table}

\newpage

\begin{acknowledgements}
The author acknowledges partial support from the OPTICON Fizeau Program, the
Heidelberger Landessternwarte, and the Air Force Research Laboratory (clearance
number 377ABW-2013-0902).  The author also acknowledges the efforts of the
anonymous referee whose comments greatly improved this work.
\end{acknowledgements}

\end{document}